\pdfoutput=1
\documentclass[acmsmall,screen,nonacm]{acmart}
\usepackage{soul}
\usepackage{listings}
\usepackage{graphicx}
\usepackage{svg,tikz,subcaption,pgfplots}
\usepackage{hyperref}
\usepackage{appendix}
\usetikzlibrary{automata, positioning, shapes, arrows}

\setcopyright{none}
\copyrightyear{2023}

\title{Automated SELinux RBAC Policy Verification Using SMT}

\author{\href{mailto:divyam.pahuja@anu.edu.au}{Divyam Pahuja}}
\affiliation{%
    \institution{The Australian National University}
    \city{Canberra}
    \country{Australia}}
\email{divyam.pahuja@anu.edu.au}

\author{\href{mailto:alvin.tang@anu.edu.au}{Alvin Tang}}
\affiliation{%
    \institution{The Australian National University}
    \city{Canberra}
    \country{Australia}}
\email{alvin.tang@anu.edu.au}

\author{\href{mailto:u7655549@anu.edu.au}{Klim Tsoutsman}}
\affiliation{%
    \institution{The Australian National University}
    \city{Canberra}
    \country{Australia}}
\email{u7655549@anu.edu.au}

\date{27 October 2023}

\begin{document}

\begin{abstract}
Security-Enhanced Linux (SELinux) is a Linux kernel module that allows for a role-based access control (RBAC) mechanism. It provides a fine-grained security framework enabling system administrators to define security policies at the system and application level. Whilst SELinux offers robust security features through a customisable, powerful RBAC model, its manual policy management is prone to error, leaving the system vulnerable to accidental misconfigurations or loopholes. We present a tool to automate the conversion of SELinux policies into satisfiability modulo theories (SMT), enabling the verification of the intended security configurations using automated theorem proving. Our tool is capable of flagging common policy misconfigurations by asserting consistency between supplied RBAC policies and the intended specification by the user in SMT. RBAC policies are inherently complicated to verify entirely. We envision that the automated tool presented here can be further extended to identify an even broader range of policy misconfigurations,  relieving the burden of managing convoluted policies on system administrators.
\end{abstract}

\ccsdesc[500]{Security and privacy~Logic and verification}

\maketitle

\section{Introduction}

To manage the permissions to access and modify files and directories, various access control models are developed and implemented in operating systems. As remarked by \citet{crypto}, access controls safeguard against accidental and spiteful threats to secrecy, authenticity and system availability. Since a computer system may be accessed by a considerable number of users, there is a risk of information falling into the wrong hands. To mitigate this risk, access control limits the activity of legitimate users in conjunction with the use of reference monitors \cite{principles}.

Typical Linux distributions utilise a discretionary access control (DAC) model, where the permissions to read, write and execute files are determined individually by their owners. Notwithstanding its power to prevent malicious parties from gaining unauthorised access \cite{dac-trojan}, the model may be inadequately sophisticated in certain fields, such as defence, medical organisations and intelligence agencies, with complex security procedures. \textbf{Security-Enhanced Linux} (SELinux), a set of kernel-level security features in the Linux operating system, provides more fine-grained access control models to achieve several cyber security principles, such as the least privilege principle and compartmentalisation of applications. Consequently, SELinux is extensively used in Android and a wide variety of other Linux distributions (\textit{e.g.}, CentOS) by default \cite{selinux-android}.

SELinux encompasses mandatory access control which employs a centrally controlled methodology of permission management. In particular, it allows for a \textbf{role-based access control} (RBAC), where the permissions are determined not only by users' identity but also their relevant security context. In this research, we focus on RBAC considering its popularity in enterprise environments that utilise hierarchies of roles amongst different individuals and processes in complex environments.

Despite a wide range of features offered by RBAC in SELinux, system administrators find it onerous to accurately maintain the policies manually. There is a risk of security vulnerabilities when the implications of policies differ from the system administrators' actual intentions. According to the Vocabulary for Event Recording and Incident Sharing (\href{https://verisframework.org/}{VERIS}) database, such access control policy misconfigurations accounted for nearly 8\% of the total data breaches in 2022. Hence, if this manual, error-prone policy verification process can be automated, it will allow for the policy misconfigurations within a system to be detected efficiently.

\section{Background}

\subsection{Context}

In modern organisations, information access should be determined by the ``need-to-know'' principle, the assessment of competency, enforcing conflict-of-interest regulations and stringent adherence to the least privilege concept \cite{rbac-nist}. Access to resources must be determined by a user's function or their role within the organisation. As such, SELinux has proven its value by encompassing a role-based access control (RBAC) design.

The RBAC implementation in SELinux operates on the principle of \textbf{type enforcement}, wherein access is granted by matching labels called \textit{types} assigned to the requesting principal and the requested resource with a predefined set of rules (policies). To accomplish role assignment, SELinux uses the concepts of \textit{users}, \textit{roles} and \textit{classes}. These attributes collectively define the \textit{security context}. With the security policies (\texttt{/etc/selinux/config}) centrally defined by the system administrator, users and processes in the system are assigned roles. Each role in turn is designated with a set of types referenced by the type enforcement mechanism.

\subsection{Related Work}

Whilst a multitude of fine-grained access control models have been in development for SELinux, researchers have also done significant work in their formalisation using logic. \citet{MLS} present a formal model of multi-level security (MLS) access control. In MLS, information is classified into different levels of sensitivity; users and processes are assigned a clearance level. The authors also include a detailed information flow analysis framework. However, they report that their model cannot be generalised to modelling RBAC and would require ``a considerable effort'' to replicate. When it comes to RBAC in SELinux, there is a lack of such formal models in the existing literature. Moreover, most organisations rely on RBAC policies within their systems, necessitating a similar modelling of RBAC in SELinux.

Following the development of a wide variety of access control models to suit different needs, researchers have historically demonstrated the models' correctness through formal logic. For instance, researchers establish a formalisation of RBAC in SELinux with description logic \cite{dickerson2006model}. \citet{rbac-isabelle} then extended this work by performing the logical analysis of these formalisations by verification using interactive theorem provers. \citet{JingBai} present analysis and verification of mandatory access control (MAC) in SELinux. The authors discussed the effectiveness and integrity analysis method for manual SELinux policy verification. Nonetheless, as noted by the authors themselves, there is a need to perform this analysis automatically. They also note that this analysis might be too complicated to achieve in complex and large-scale systems. We observe this gap several times in the literature. \citet{SELAC} conclude similar results for their policy verification tool SELAC. In their research, they also highlight the benefits of using automated theorem provers.

Our review of the current state-of-the-art also reveals that there has been limited existing work regarding the use of automated theorem provers to help SELinux users configure security policies correctly. Some research, despite such attempts, does not present a comprehensive tool. For instance, the description logic model proposed by \citet{dickerson2006model} does not consider some significant SELinux policy keywords (\textit{e.g.}, \texttt{neverallow}, \texttt{role\_transition}) justified by its purpose of allowing queries to extract useful data from the configurations about running programs. Automated theorem provers, such as SMT solvers, may substantially alleviate the burden on system administrators to ensure systems' adherence to intended security protocols \cite{automated-prover}. In this paper, we therefore present a tool to automate the process of policy verification by formalising the RBAC policies into satisfiability modulo theories (SMT) statements \cite{HandbookSatisfiability}.

\subsection{Methodology}

We present a tool utilising SMT to detect policy misconfigurations. Our research focuses on mitigating the risk of security vulnerabilities when the implications of policies differ from system administrators' actual intentions. We address this concern by providing insights on the following research questions.
\begin{enumerate}
    \item How can the RBAC and type enforcement design be formalised using logic?
    \item How can automated theorem proving contribute to the detection of policy misconfigurations in practice?
    \item How is the suitability of automated theorem proving for policy verification in terms of its scalability and efficiency?
\end{enumerate}

In particular, our focus is on the most common misconfigurations in RBAC --- over-permissive or over-restrictive access control with a combination of \texttt{allow} and \texttt{neverallow} policies.

SMT is a powerful automated reasoning technique with its ability to determine whether given logical formulae are satisfiable. The choice of SMT is prompted by its support for first-order logic, which establishes the foundation for the representation of role-type relationships, as well as its scalability to handle a large number of logic statements. Due to the expressiveness of SMT, the reasoning of the inherently contextual RBAC policies in SELinux is easily modelled. It further supports incremental solving, that is, in the case of a policy change, new constraints corresponding to the change could be incrementally added to the solver’s existing constraints, enabling the extensibility of our work in the future.

Upon formalising the basic components of RBAC in SELinux using SMT, we proceed to present our tool that automates the conversion of SELinux policy into SMT statements with syntax parsing, followed by the verification process with the Z3 SMT solver.

\section{Formalising RBAC in SMT}
\label{sec:formalisation}

In this section, we present the logic formalisation of the RBAC as SMT statements according to the syntax specified in \citet{smt-lib}.

\subsection{Defining Context Elements}
\label{sec:context}

The context elements in RBAC are users, roles, types, classes and permissions. Their discrete nature can be represented by \textit{sorts} in SMT to model categorical components of the RBAC design. These \textit{sorts} can then be integrated with functions in the SMT solver to form access vector rules (\textit{i.e.}, the actual statements forming security policies). The concept of a user is declared as a sort in the theorem prover with \texttt{(declare-sort User)}. We can then declare a constant of the corresponding sort in SMT to denote the existence of a new context element in the SELinux policies. For instance, a user named \texttt{root\_u}, which generally corresponds to the root user, is formalised as \texttt{(declare-const root\_u User)}. For each context element, we correlate the object with its unique identifier (UID). With the function declared with \texttt{(declare-fun user-id (User) Int)}, we can define the UID of a given user. We then use it using an assertion statement. For example, \texttt{(assert = (user-id root\_u) 0)} assigns a unique identifier of \texttt{0} to \texttt{root\_u}.

In Linux distributions targeting enterprises (\textit{e.g.}, CentOS), SELinux is often embedded to enable additional security measures. A standard user in Linux is given a separate identity in SELinux. Multiple Linux users, however, can possess the same SELinux user identity. In the context of this SMT formalisation, User objects refer to the SELinux user identity. 

Note that the context elements in SELinux might share the same name. However, the SMT solver prohibits the coexistence of multiple constants sharing the same name. Hence, to counter this, we develop a standardised convention to name the constants for signifying their element nature. A suffix of ``\texttt{\_u}'', ``\texttt{\_r}'' and ``\texttt{\_t}'' shall correspond to users, roles and types. Meanwhile, keywords including ``\texttt{file}'', ``\texttt{dir}'' and ``\texttt{socket}'' shall be reserved as classes. The SMT solver will detect a contradiction between two context elements of the same category illegally sharing the same identifier with such assertions as follows.
\begin{verbatim}
    (assert (not (exists ((u1 User) (u2 User)) (= (user-id u1) (user-id u2)))))
\end{verbatim}

The above sorts and assertions establish the foundation of the logic formalisation in the SMT solver so these statements are always included in the theorem prover, regardless of the actual SELinux policy configurations.
Subsequently, we define predicates to establish a relationship between these three context elements.

\subsection{Defining Context Relationships}

After defining users, roles, types and classes, we can specify the relationships between them. In particular, we define the assignment of roles to users, as well as the memberships of types in roles. This will be modelled with functions namely \texttt{user-has-role} and \texttt{role-has-type} in the SMT solver. As an illustration, the statement \texttt{(assert (user-has-role system\_u sysadm\_r))} indicates that the user \texttt{system\_u} is assigned with the role \texttt{sysadm\_r}.

Additionally, SELinux has introduced the concepts of \textit{attribute roles} and \textit{attributes}, which represent a set of roles and a set of types respectively. These allow security administrators to categorise individuals and system processes into groups. If a role \texttt{sysadm\_r} belongs to an attribute role \texttt{role\_list\_1} which is assigned with the type \texttt{tmp\_t}, then \texttt{sysadm\_r} is assigned with the type \texttt{tmp\_t}. This is modelled in the theorem prover with the functions named \texttt{role-has-attribute-role} and \texttt{type-has-attribute}. We utilise these functions to transform policies with \texttt{roleattribute} and \texttt{typeattribute} keywords in SELinux into statements in the theorem prover. These functions are used to form hierarchies of roles and types, which are ultimately used for realising the security policies using \textit{access vector} (AV) rules.

\subsection{Defining Access Vector Rules}

Since the actual permissions to perform actions to system resources are managed by type enforcement in SELinux, AV rules are the most critical components defining privileges. The automated theorem prover contains predicates to indicate whether a user has permission to perform certain actions. In our implementation, we focus on \texttt{allow} and \texttt{neverallow} access vectors. The \texttt{dontaudit} and \texttt{auditallow} rules can be taken into account analogously. 

With reference to the SELinux documentation \cite{selinuxproject-avc}, all these access vector rules are structured in the format of ``\texttt{rule\_name source\_type target\_type : class perm\_set;}''. In the SMT solver, this is modelled by predicates indicating whether the access vector rules exist in the security policy configuration.

We formalise the two access vector rules as predicates named \texttt{av-allow} and \texttt{av-neverallow} respectively. They take in four inputs, including a source type (the type assigned to the principal), a target type (the type assigned to the resource), the class and the requested permission. We also take into account that \texttt{neverallow} AV rules supersede \texttt{allow} rules by the design of the SELinux kernel.

\bigskip

With the above formalisations, we can now model the key SELinux policies into a set of SMT statements. For example, if system administrator would like users with \texttt{sysadm\_t} type to read a file assigned with \texttt{shadow\_t}, we can model the policy ``\texttt{allow sysadm\_t shadow\_t : file read;}'' into SMT as \texttt{(assert (av-allow sysadm\_t shadow\_t file read))}. Such translation processes are, however, repetitive and prone to human error if done manually. Therefore, in our implementation of the policy verification tool, we not only apply the automated theorem prover but also automate the translation of SELinux policies into SMT statements with a parser.

\section{Implementation}

The entire automated verification process with \href{https://github.com/tsoutsman/comp2560}{our tool} is completed in three major steps.
\begin{enumerate}
    \item Parse and translate the SELinux policy configuration file provided by the user.
    \item Accept security specifications as additional constraints in SMT from the user.
    \item Use Z3 as the automated theorem prover for satisfiability check to determine whether the policies contradict the specifications.
\end{enumerate}

\subsection{Parsing and Translation of SELinux Policies into SMT Statements}
\label{sec:parsing}

In the first step of the verification process, the built-in \texttt{checkpolicy} tool in SELinux and the \texttt{libespol} library are used in our parser. The \texttt{checkpolicy} tool, with its main purpose of compiling human-readable SELinux configurations into binary format, identifies syntactical errors in the policies during the process. Its integration with Yet Another Compiler-Compiler (YACC), which helps build powerful, tailored parsers, facilitates the translation of provided policies into SMT. Functions in \texttt{libespol} help extract the SELinux keywords, such as \texttt{roleattribute} and \texttt{allow}, in policy statements, assisting us in choosing the corresponding SMT functions as described in \autoref{sec:formalisation}.

Tools and libraries for SELinux policy analysis, including \texttt{checkpolicy} and \texttt{libsepol}, are written in the C programming language. We chose Rust for the parsing and translation process due to its mature ecosystem for building and linking to C programs. Unlike C, Rust allows for automatic and efficient memory management as well as software security protection. The use of Rust, thus, helps in abstracting tedious low-level programming tasks with C. \cite{rustC}

\begin{figure}[ht]
    \centering
    \vspace{-0.3cm}
    \begin{small}
    \begin{tikzpicture}
        [node distance=2.6cm, node/.style={draw, text centered, minimum height=0.6cm, rounded corners, fill=blue!20}, >=stealth]
        \fill[yellow!40!white, rounded corners] (0.95,0.05) rectangle (10.95,3.8);
        \node at (2,0.5) {\large{\textbf{Tool}}};

        \node[node] at (0,2) (user) {User};
        \node[node, right of=user, yshift=1cm] (parseutil) {\texttt{parse\_util.c}};
        \node[node, right of=parseutil, xshift=1.15cm] (modelrs) {\texttt{model.rs}};
        \node[node, right of=modelrs, xshift=0cm, yshift=-1cm] (pyrs) {\texttt{py.rs}};
        \node[node, right of=modelrs, xshift=0cm] (libsepol) {\texttt{libsepol}};
        \node[node, below of=pyrs, yshift=1.5cm] (concat) {Concatenator};
        \node[node, left of=concat, xshift=-2.2cm] (const) {Constant};
        \node[node, right of=concat, node distance=2.6cm] (python) {Python};
        \node[node, right of=python, node distance=1.4cm] (z3) {Z3};

        \draw[->] (user) edge[bend left=10] node[near start, xshift=-0.2cm, above, text width=2cm, align=left] {Policy file} (parseutil);
        \path[->] (parseutil) edge node[midway, above, align=center] {C policy\\struct} (modelrs);
        \draw (modelrs) edge[bend left=16,->] node[above] {} (libsepol);
        \draw (libsepol) edge[bend left=16,->] node[above] {} (modelrs);
        \draw (modelrs) edge[midway, bend right=8, align=center, ->] node[left, yshift=-0.2cm] {Rust policy\\struct} (pyrs);
        \draw (pyrs) edge[bend left=20, ->] node[midway, right, align=left] {Python code} (concat);
        \path[->] (concat) edge node[midway, below, text width=1cm, align=center] {Python code} (python);
        \draw (user) edge[bend right=2,->] node[midway, above, yshift=0.1cm, text width=2cm] {Additional constraints} (concat);
        \path[->] (python) edge [bend left=30] (z3);
        \path[->] (z3) edge [bend left=30] (python);
        \path[->] (const) edge node[midway, below, text width=4cm, align=center] {Logic\\formalisation model} (concat);
    \end{tikzpicture}
    \end{small}
    \vspace{-0.3cm}
    \caption{Overall Flow Structure of the Automation}
    \label{fig-impl}
\end{figure}
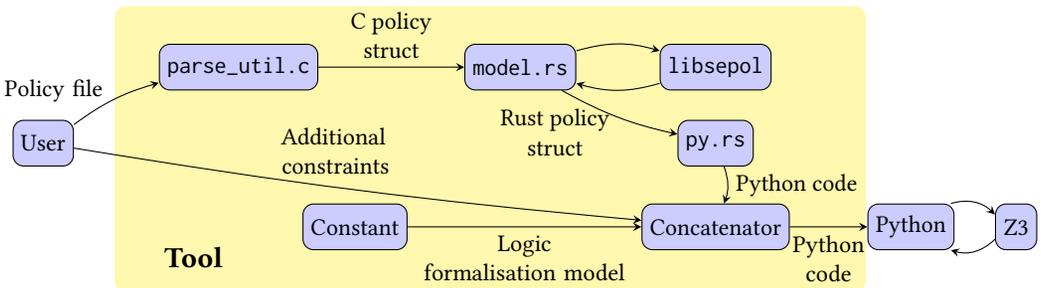

The entire verification process workflow is presented in \autoref{fig-impl} with the following files as the main components.
\begin{itemize}
    \item \texttt{src/c}: tools from the original SELinux repository, including \texttt{checkpolicy} and \texttt{libespol}.
    \item \texttt{src/model.rs}: representation of the context elements in \autoref{sec:formalisation} as structs.
    \item \texttt{src/py.rs}: conversion of the structs into SMT (Z3) statements in the Python language.
    \item \texttt{additional.py}: additional constraints provided by the user.
\end{itemize}

When the raw SELinux configuration file (usually as \texttt{/etc/selinux/config}) is input into the tool, the \texttt{read\_source\_policy} function in \texttt{parse\_util.c}, which is a program file in \texttt{src/c}, returns a \texttt{policydb} object defined in \texttt{libsepol} representing the policy file. We then use functions from \texttt{libsepol} to convert the C representation into Rust. The Rust policy struct design handles only policies accounted for in \autoref{sec:formalisation}, gatekeeping invalid policies from being converted into Python code. The tool subsequently concatenates the foundational logic statements presented in \autoref{appendix:formalisation}, Python code generated by \texttt{py.rs} and additional constraints provided by the user in \texttt{additional.py}. The resultant Python code can be directly run with Z3, the automated theorem prover.

\subsection{Automated Theorem Proving with Z3}

Compiled SELinux policies are often thousands of lines long, so the theorem prover performance is essential. Z3 is generally regarded as the most mature and performant prover. Once the parsing and translation are complete, the user obtains a complete, executable Python script that relies on the Python API of Z3. 

The resultant output from Z3 is straightforward --- either ``\texttt{sat}'' or ``\texttt{unsat}''. This indicates whether the SMT statements are satisfiable. An ``\texttt{unsat}'' result implies that there are contradictions in the logic statements. Since SELinux policies cannot inherently conflict with each other, the SMT statements generated by \texttt{py.rs} are always semantically sound. Even if there exists \texttt{allow} and \texttt{neverallow} access vector rules defined for the same set of security context, SELinux has defined that \texttt{neverallow} policies supersede \texttt{allow} AV rules. Therefore, the only cause of unsatisfiability is the contradiction between policies from \texttt{py.rs} and user-provided constraints from \texttt{additional.py}.

Ultimately, this accessible and convenient tool enables system administrators to gradually develop SELinux policies from scratch easily when new systems are set up. Before establishing new policies, \texttt{additional.py} can be first formulated as the requirements. As the RBAC policies are being created, misconfigurations can be easily ruled out, reducing the time and effort required for debugging. For modification of existing security configurations, this tool is equally useful to prevent changes from violating system administrators' intended security requirements.

\section{Evaluation}

To demonstrate \href{https://github.com/tsoutsman/comp2560}{our tool}'s practicality, efficiency and scalability, we hereby present its evaluation. Firstly, we showcase its effective application in a real-life policy misconfiguration scenario, illustrating its need in the industry. We then analyse its performance using synthetic benchmarks to ascertain its scalability.

\subsection{Case Study: Policy Misconfiguration in Android 11}
\label{sec:case-study}

A real-life example of policy misconfiguration is CVE-2021-0691 in Android 11 identified in 2021 \cite{CVE-2021-0691}. Android 11 has SELinux enabled by default. Developers of Android at Google defined the access control policies in a file named \texttt{system\_app.te}. The file refers to type enforcement for \texttt{system\_app}, an SELinux type for system applications. This type is widely used by pre-installed applications on Android mobile phones, such as phone calling and SMS messaging.

These Android system applications are fundamental for the basic functionality of mobile phones. They are granted more permissions than normal apps as their functionalities may require modification of system settings. However, unlike the system server in Android, system apps require system-level permissions for low-level tasks. Therefore, in order to define concisely the permissions granted to system apps, the SELinux policies have to be very fine-grained with type enforcement for \texttt{system\_app}.

The problem identified in CVE-2021-0691 is that the policies in \texttt{system\_app.te} are overly permissive, offering local escalation of privilege with system execution privileges. In particular, the following policies are included in the configuration file according to the source code of the Android 11 system \cite{android11}.
\begin{verbatim}
    allow system_app apk_data_file : file write;
    allow system_app incremental_control_file : file r_file_perms;
\end{verbatim}

These two policies enable system apps to write data to \texttt{/data/incremental} and read the relevant logs. \texttt{/data/incremental} is the directory for Over-The-Air (OTA) incremental updates. It is intended for updates of applications to the latest version. Having these two over-permissive policies, malicious parties may take advantage of system apps to create APK files in \texttt{/data/incremental}. Since the Android system installs updates in this directory automatically and regularly, adversaries could then easily install malicious apps to read and write data stored on victims' phones.

Considering that there are over 200 type enforcement policy configuration files, each with around 100--200 lines of policies, even dedicated security analysts would be unable to fully guarantee the correctness, especially when the Android system source code is constantly being developed and expanded. On the other hand, our tool can be utilised to largely automate this verification process in such situations as follows.

Note that the source type and the target type in the first policy are \texttt{system\_app} and \texttt{apk\_data\_file} respectively. Subject to the naming convention mentioned in \autoref{sec:context}, our tool parses them as SMT constants \texttt{system\_app\_t} and \texttt{apk\_data\_file\_t}, which are of the \texttt{Type} sort. The policy states that principals with the \texttt{system\_app} type may write files to resources assigned with the type \texttt{apk\_data\_file}. Therefore, the tool's parser and translator generate the SMT statement \texttt{(assert (av\_allow system\_app\_t apk\_data\_file\_t file write))}.

Meanwhile, it is clear to Android developers and security analysts who should be able to write files to \texttt{/data/incremental}. They supply the tool with corresponding constraints in SMT. To be precise, there are only a few principals that can write files to the directory, including \texttt{installd} and \texttt{recovery} types. The former corresponds to the Android system installer that installs and removes apps. The latter is for the Android recovery mode.

To uphold the least privileges principle, access control policies should be defined such that no one should have permission to certain resources unless the system administrator explicitly allows it. In the case of managing permissions to \texttt{/data/incremental}, system administrators using our tool can specify the security specifications in \texttt{additional.py} as described in \autoref{sec:parsing}. For example, they can create an attribute named \texttt{adb\_data\_file\_list} where only, say, \texttt{installd} and \texttt{recovery} have membership therein.

The assignment of the attribute is then parsed by our tool into the following statements.
\begin{verbatim}
    (assert (declare-const adb_data_file_list Type))
    (assert (type-has-attribute installd_t adb_data_file_list))
    (assert (type-has-attribute recovery_t adb_data_file_list))
\end{verbatim}

To specify that no other target types other than those in \texttt{adb\_data\_file\_list} should be granted permission, the system administrator may provide the additional constraint as follows.
\begin{verbatim}
    (assert (forall (t Type) (=> (type_has_attribute t adb_data_file_list)
                                 (not (av-allow t adb_data_file_t file write))))
\end{verbatim}

From the over-permissive SELinux policy statement for \texttt{system\_app\_t}, our parser produces an SMT statement that states \texttt{system\_app} is allowed this permission. However, since \texttt{system\_app} is not a member of \texttt{apk\_data\_file\_list}, the parser-generated SMT statement contradicts the system administrator-provided additional constraint. The SMT statements generated by our tool is in \autoref{appendix:android11}. As a result, in the automated theorem proving step with Z3, an ``\texttt{unsat}'' output is obtained, correctly identifying the vulnerability.

This vulnerability was identified by Google and publicised to the US National Vulnerability Database in June 2021. It was not until September 2021 that Google released a security patch to fix the problem. If our automated policy verification tool was used to ensure security compliance, such vulnerabilities could be easily detected and fixed timely, or even be avoided prior to deployment.

\subsection{Scalability and Efficiency: Synthetic Benchmarking}

In real-world implementations of RBAC with SELinux, there are numerous users, processes and resources managed in the system. Therefore, system administrators use a large number of context elements to take advantage of fine-grained access control. The automated parsing and verification tool must be efficient enough such that a verification result can be obtained within a reasonable amount of time.

We conducted synthetic benchmarking to provide a quantitative measure of the scalability of the system. The benchmark consisted of $n$ classes, $n$ roles, $n$ users, $n$ rules, $2n$ types and 1 satisfiable constraint. Each class had 3 permissions and each rule $i$ allowed all users with role $i$ all three permissions for class $i$ with type $i$. An example of the benchmark for $n = 2$ is presented in \autoref{appendix:benchmark}. We ran the benchmark on an M2 10-core Apple MacBook Pro with 16GB of memory to imitate a system administrator's machine.

Note that we did not use multi-threading, and memory consumption by the process was not a bottleneck.

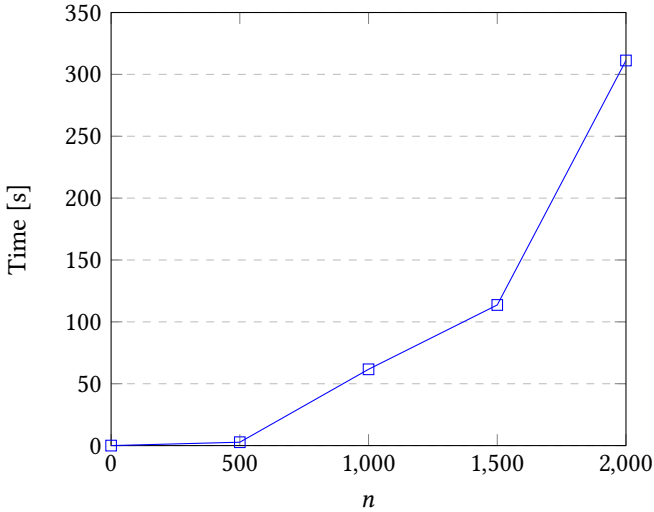
\begin{figure}[ht]
    \centering
    \begin{tikzpicture}
    \begin{axis}[
        title={},
        xlabel={$n$},
        ylabel={Time [s]},
        xmin=0, xmax=2000,
        ymin=0, ymax=350,
        xtick={0,500,1000,1500,2000},
        ytick={0,50,100,150,200,250,300,350},
        legend pos=north west,
        ymajorgrids=true,
        grid style=dashed,
    ]

    \addplot[
        color=blue,
        mark=square,
        ]
        coordinates {
        (0,0)(500,2.73)(1000,61.65)(1500,113.65)(2000,311.23)
        };

    \end{axis}
    \end{tikzpicture}
    \caption{Benchmarking Results}
\end{figure}

Even for $n = 2000$, the C and Rust portion of the tool took less than 100 milliseconds to run --- the bottleneck was Z3. However, we believe that a runtime of 5 minutes is acceptable for a policy with 12000 SELinux statements. Although synthetic benchmarks are not necessarily indicative of real-world results, we believe that the sheer size of the benchmarks used proves that our tool is sufficiently scalable.

Our tool could be used in continuous integration (CI) to decrease maintenance burden by ensuring any modifications do not invalidate the specified constraints. Continuous integration often uses much more powerful hardware which would decrease the runtime of our tool. Additionally, the runtime in CI is much less important, with large projects often having CI runtime in excess of 30 minutes \cite{rust-ci}.

\section{Conclusion}

In this paper, we present a novel tool for automating the verification of SELinux policies using the Z3 SMT solver. Our main contributions, in addition to the automation of policy verification itself, include a formalisation of SELinux RBAC policies in SMT. We detail and establish a systematic approach to semantically model SELinux policies as logical formulae, allowing the translation of SELinux policies to SMT statements. Currently, our tool is specialised to detect over-permissive and over-restrictive RBAC policy misconfigurations only. However, our formalisation and implementation lay the groundwork for a more comprehensive tool that can extend our current tool to identify an even broader range of policy misconfigurations.

Further work could use Z3's unsatisfiability core feature to reduce the offending policies to the smallest invalid subset, which helps locate the problem more accurately in the original policy files. This could be aided by tagging SMT statements with policy file line numbers from which those statements are generated. The translation of SELinux policies to SMT statements opens up doors for a plethora of extensions. For example, we envision that our tool can be made further comprehensive by incorporating an attack surface analyser, also possibly suggesting optimisations by coalescing duplicate context elements. This would allow system administrators to not only verify their policy formulations but also foresee and fix any loopholes that might be present in their current policy structure. Our tool currently only accepts any additional constraints to the SELinux policy from the user in SMT statements. This could be extended to accept the entire intended policy behaviour in SMT and output the actual, pre-verified policy file. We also foresee a feasible extension to reverse translate SMT statements to SELinux policies.

\newpage

\begin{acks}
    We would like to express our gratitude to AsPr Alwen Tiu for providing continuous support for this project as our supervisor. We greatly appreciate the guidance to not only ensure success in our work but also consolidate our practical experience in the fields of cyber security and formal logic.
\end{acks}

\bibliographystyle{ACM-Reference-Format}
\bibliography{references.bib}

\newpage

\begin{appendices}

\section{Foundational SMT Statements}
\label{appendix:formalisation}

The following SMT statements form the foundation of the formalisation model. They are always included as the prelude in the SMT solver input. The equivalent Python representation is available in \href{https://github.com/tsoutsman/comp2560/blob/main/parse-policy/src/py.rs}{\texttt{parse-policy/src/py.rs}} in the implementation.

\bigskip

\begin{small}
\begin{verbatim}
    ; Declare context elements
    (declare-sort User)
    (declare-sort Role)
    (declare-sort Type)
    (declare-sort Class)
    (declare-sort Permission)

    ; Declare functions for unique identifiers
    (declare-fun user-id (User) Int)
    (declare-fun role-id (Role) Int)
    (declare-fun type-id (Type) Int)

    ; Assert that context elements do not share the same ID
    (assert (not (exists ((u1 User) (u2 User)) (= (user-id u1) (user-id u2)))))
    (assert (not (exists ((r1 Role) (r2 Role)) (= (role-id r1) (role-id r2)))))
    (assert (not (exists ((t1 Type) (t2 Type)) (= (type-id t1) (type-id t2)))))

    ; Declare functions for context relationships
    (declare-fun user-has-role (User Role) Bool)
    (declare-fun role-has-type (Role Type) Bool)
    (declare-fun type-has-attribute (Type Type) Bool)
    (declare-fun role-has-attribute-role) (Role Role) Bool)

    ; Define access vector rules as predicates
    (declare-fun av-allow (Type Type Class Permission) Bool)
    (declare-fun av-neverallow (Type Type Class Permission) Bool)

    ; Define functions for context relationships
    (assert (forall ((r1 Role) (r2 Role) (t Type)))
        (=> (and (role-has-type r2 t) (role-has-attribute-role r1 r2))
            (role-has-type r1 t)))
    (assert (forall ((t1 Type) (t2 Type) (tt Type) (c Class) (p Permission)))
        (=> (and (type-has-attribute t1 t2) (av-allow t2 tt c p))
            (av-allow (t1 tt c p))))
    (assert (forall ((st Type) (t1 Type) (t2 Type) (c Class) (p Permission)))
        (=> (and (type-has-attribute t1 t2) (av-allow st t2 c p))
            (av-allow (st t1 c p))))

    ; Define neverallow AV rules supersede allow AV rules
    (assert (forall ((st Type) (tt Type) (c Class) (p Permission)))
        (=> (and (av-allow st tt c p) (av-neverallow st tt c p))
            (not (av-allow st tt c p))))
\end{verbatim}
\end{small}

\newpage

\section{SMT Statements for the Android 11 Case Study}
\label{appendix:android11}

Our tool parses the policies for the case study in \autoref{sec:case-study} and generates the following statements.

\begin{small}
\begin{verbatim}
    ; Declare the types for system resources
    (assert (declare-const system_app_t Type))
    (assert (declare-const installd_t Type))
    (assert (declare-const recovery_t Type))

    ; Declare the type attribute
    (assert (declare-const adb_data_file_list Type))

    ; Define the access vector rule
    (assert (av-allow adb_data_file_list apk_data_file_t file write))

    ; Define the membership of types in the type attribute
    (assert (type-has-attribute installd_t adb_data_file_list))
    (assert (type-has-attribute recovery_t adb_data_file_list))
\end{verbatim}
\end{small}

\bigskip

\noindent The additional constraint is supplied to our tool as the final part of the SMT statements.

\begin{small}
\begin{verbatim}
    (assert (forall (t Type) (=> (type_has_attribute t adb_data_file_list)
                                 (not (av-allow t adb-data_file_t file write)))))
\end{verbatim}
\end{small}

\newpage

\section{Synthetic Benchmark}
\label{appendix:benchmark}

The following is an SELinux policy example of the benchmark generated for $n = 2$.

\bigskip

\begin{small}
\begin{verbatim}
class class_0
class class_1

sid unlabeled

class class_0
{
    class_0_perm_0
    class_0_perm_1
    class_0_perm_2
}
class class_1
{
    class_1_perm_0
    class_1_perm_1
    class_1_perm_2
}

role role_0;
type role_type_0;
role role_0 types role_type_0;
role role_1;
type role_type_1;
role role_1 types role_type_1;

type type_0;
type type_1;

allow role_type_0 type_0 : class_0 { class_0_perm_0 class_0_perm_1 class_0_perm_2 };
allow role_type_1 type_1 : class_1 { class_1_perm_0 class_1_perm_1 class_1_perm_2 };

user user_0 roles { role_0 };
user user_1 roles { role_1 };

sid unlabeled user_0:object_r:type_0
\end{verbatim}
\end{small}

\newpage

\noindent The benchmark is tested with the following constraints in Python.

\bigskip

\begin{small}
\begin{verbatim}
constraint_0_u = Const("constraint_0_u", user)
constraint_0_r = Const("constraint_0_r", role)
constraint_0_t = Const("constraint_0_t", type)

solver.add(
    ForAll(
        [constraint_0_u],
        Implies(
            user_has_role(constraint_0_u, role_0),
            ForAll(
                [constraint_0_r],
                Implies(
                    user_has_role(constraint_0_u, constraint_0_r),
                    ForAll(
                        [constraint_0_t],
                        Implies(
                            role_has_type(constraint_0_r, constraint_0_t),
                            av_allow(constraint_0_t, type_0, class_0, class_0_perm_0)
                                == True,
                        ),
                    ),
                ),
            ),
        ),
    )
)
\end{verbatim}
\end{small}

\end{appendices}

\end{document}